\documentclass[conference]{IEEEtran}
\usepackage{amsmath}
\usepackage{amssymb}
\usepackage{amsthm}
\usepackage{amsfonts}
\usepackage{graphicx}
\usepackage{epsfig}
\usepackage{subfigure}
\usepackage{psfrag}
\usepackage{epstopdf}
\usepackage{graphicx,color,psfrag,subeqnarray}
\usepackage{cite}
\usepackage{latexsym}
\usepackage{url}
\usepackage{color}
\usepackage{bigstrut}
\usepackage{multirow}
\usepackage{authblk}
\usepackage{bm,array}
\usepackage{array}
\DeclareGraphicsExtensions{.pdf,.jpeg,.png,.jpg}
\usepackage{lipsum}
\usepackage{authblk}
\begin{document}
\title{Shared Energy Storage Management for Renewable Energy Integration in Smart Grid}
\author[1]{Katayoun Rahbar\vspace{-2mm}}
\author[2]{Mohammad R. Vedady Moghadam}
\author[1,2]{Sanjib Kumar Panda}
\author[1]{Thomas Reindl}
\affil[1]{Solar Energy Research Institute of Singapore, Singapore}
\affil[2]{ECE Department, National University of Singapore, Singapore}
\affil[ ] {E-mail: \{serkr, elemrvm, eleskp, sertgr\}@nus.edu.sg}
\maketitle
\thispagestyle{empty}
\begin{abstract}
Energy storage systems (ESSs) are essential components of the future smart grid to smooth out the fluctuating output of renewable energy generators. However, installing large number of ESSs for individual energy consumers  may not be practically implementable, due to  both the space limitation and high investment cost.
As a result, in this paper, we study the energy management problem of multiple users with renewable energy sources and  a single {\it shared ESS}.  To solve this problem, we propose an algorithm that  jointly optimizes the energy charged/discharged to/from the shared ESS given a profit coefficient set that specifies the desired proportion of the total profit allocated to each user, subject to practical constraints  of the system. We 
conduct simulations based on the real  data from California, US, and show that the shared ESS can potentially increase the total profit of all users by $10\%$ over the case that users own individual  small-scale ESSs with no energy sharing. %, which is a remarkable improvement. 
%Our simulations show that the shared ESS can potentially increase the total profit of users (given the capacity of the shared ESS to be the same as the sum capacity of all individual ESSs), since surplus energy of one user can be utilized by another one with energy deficit and also avoid the energy curtailment more effectively due to its higher capacity. Moreover, we show that the diversity in users' renewable energy integration can highly increase  the total profit gain resulting from the shared ESS in the system, i.e., users with different renewable energy sources is better to deploy a shared ESS. 
\end{abstract}
\newtheorem{definition}{\underline{Definition}}[section]
\newtheorem{fact}{Fact}
\newtheorem{assumption}{Assumption}
\newtheorem{theorem}{\underline{Theorem}}[section]
\newtheorem{lemma}{\underline{Lemma}}[section]
\newtheorem{corollary}{\underline{Corollary}}[section]
\newtheorem{proposition}{\underline{Proposition}}[section]
\newtheorem{example}{\underline{Example}}[section]
\newtheorem{remark}{\underline{Remark}}[section]
\newtheorem{algorithm}{\underline{Algorithm}}[section]
\newcommand{\mv}[1]{\mbox{\boldmath{$ #1 $}}}
\section{Introduction}
The fast-growing electric energy consumption has become a serious concern for existing power systems. According to the study reported by the US energy information administration (EIA), the worldwide energy consumption will grow by $56\%$ from 2010 to 2040 \cite{Demand}. This motivates a green power system with users widely deploying distributed renewable energy generators to meet their individual demand locally, which can effectively reduce both the carbon dioxide emissions of traditional fossil fuel based power plants and the  transmission losses from power plants to far apart users. However, the intermittent and stochastic characteristics of renewable energy sources can cause imbalanced supply with demand and yield fluctuation in the power system frequency and/or voltage \cite{Vedady}. 

Deploying energy storage system (ESS) is a practical solution to smooth out the power fluctuation in the renewable energy generation and improve the system reliability \cite{Palomar,Katayoun_SmartGridComm2014,Katayoun_journal}. In practice, integrating individual ESSs for all energy consumers (especially for residential and commercial users) may not be feasible, due to  both the space limitation and high capital cost of the large number of ESSs. 
With the technology advances in bidirectional power flow and distributed monitoring and control in smart grids,  the concept of {\it shared ESS} has become appealing \cite{Z.Wang,Paridari,W.Tushar}. In this case, the surplus renewable energy of some users can be charged into a shared (common) ESS, and  then be discharged by others with renewable energy deficit. Moreover, given the real-time/day-ahead price information,  the shared ESS can  reduce the total cost of purchasing conventional energy from the main grid  by being charged  within off-peak-demand period, with low electricity prices, and being discharged during peak-demand period, with high electricity prices \cite{Z.Wang}. However, the main challenge to realize the shared ESS  is how to manage users to optimally charge/discharge energy to/from the shared ESS, which will be addressed in this paper.

In this paper, we  consider a system of multiple energy consumers with their individually owned renewable energy generators, and one ESS shared among them. Since users are self-interested in practice, each of which  wants to use the shared ESS as much as possible to maximize its profit, i.e., the energy cost saving resulted from deploying the shared ESS. This can cause fairness issue in general, i.e., one user may use the shared ESS more frequently than others. To tackle this issue, we propose a centralized algorithm, under which a central controller jointly optimizes the amount of energy charged/discharged to/from the shared ESS by all users, given a profit coefficient set that specifies the desired proportion of the total profit allocated to each user.
%we  apply the well-known  technique of profit coefficients, where  a central controller adjusts the proportion of the total profit received by individual users by  jointly optimizing the amount of energy charged/discharged to/from the shared ESS by all users. 
Next, for performance comparison, we formulate the profit maximization problem for the conventional case of distributed ESSs, where each user owns its small-scale ESS and does not share energy with others. To have a fair comparison, we set  the sum capacity of all individual ESSs to be the same as the shared ESS. Our simulations show that the shared ESS can potentially increase the total profit of users compared to the case of distributed ESSs. This is because the surplus energy of one user can be utilized by others with energy deficit, and also the energy curtailment is  avoided more effectively due to the higher capacity of shared ESS. Moreover, we show that the diversity in users' renewable energy sources/loads can highly increase  the  profit gain resulting from the shared ESS, i.e., users with different types of renewable energy sources can benefit  significantly by sharing an ESS. 

There have been previous works on the energy management problem for users with ESSs \cite{Palomar,Katayoun_SmartGridComm2014,Katayoun_journal,Paridari,W.Tushar,Z.Wang}. 
Most of the previous works, e.g., \cite{Katayoun_journal,Katayoun_SmartGridComm2014,Palomar}, assume that either each user owns its ESS (that is not shared with others) or all users belong to the same entity, where maximizing the profit of each individual user is not the case. However, this might not always be valid in practical systems, especially when the number of users is large. The idea of shared ESS among users was introduced in \cite{Z.Wang}, where interesting results on the shared ESS deployment in the system were presented. However, the proposed charging/discharging policy in \cite{Z.Wang} is only based on the hourly prices offered by the main grid. Recently,  \cite{Paridari} solved the cost minimization problem for energy consumers with demand response capability, but no renewable energy integration, which limits the application of their derived solution. Moreover, \cite{W.Tushar} proposed an auction based approach for the interactions between the shared ESS and users,  using game theoretical techniques. 
%; as a result, it does not make decision on the proportion of  ESS used by individual users. \textcolor{red}{[To be completed]}. 
%game theoretical approaches, we choose another approachh. Our approach provide insights into the problem.

%[] and [] also consider ...
%In this paper, we mainly price based algorithms. We want to make comparison with the case of distributed . the effectiveness of shared not shown

In contrast to the prior works, in this paper, we propose an algorithm under which a central controller optimally sets the charging/discharging power to/from the shared ESS by all users, using  the profit coefficients technique. We then compare the profit gain in this case with the extensively studied case of users with distributed small-scale ESSs. The rest of the paper is organized as follows. Section \ref{Sec:System_Model} describes the  system model. Section \ref{Sec:Shared_ESS} formulates and solves the energy management problem for the shared ESS. %This enables the system operator to  allocate the shared ESS consumption for each user based on a specific objective, e.g., their renewable energy integration or the money each user invested for the purchase of the shared ESS. 
Section \ref{Sec:Benchmark} describes the case of  
distributed  ESSs. Section \ref{Sec:Simulations} presents our simulation results. %, which reveal  that  the shared ESS %that first studied due to the constraints on the required space for installment and the high cost (especially when number of users is large) 
%can  help increasing the total profit of all users over the case of distributed small-scale ESSs. 
%We also discuss cases where the shared ESS can result in significant cost reduction  Our results are helpful to further integrate renewable  energy sources Our results show the practicality of further integrating renewable energy sources in the system especially building level.
Last, Section \ref{Sec:Conclusion} concludes the paper and discusses possible future directions.
%Our results are helpful in practical systems where individual ESS for each user is either very costly (due to the large number of users) or requires space that is not available. 
%**********************************
\section{System Model}\label{Sec:System_Model}
\begin{figure}[t!]
\centering
\includegraphics[width=7.6cm]{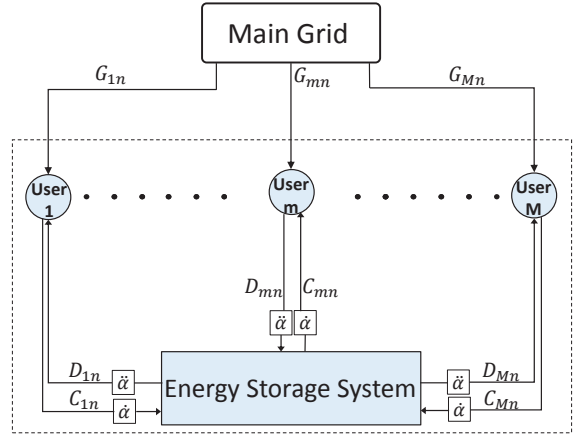}\\ 
\caption{Case of shared ESS.}\label{fig:SystemModel}\vspace{-4mm}
\end{figure}

We consider a time-slotted system with slot index $n$, $n \in {\cal N}=\{1,\ldots,N\}$, where $N \ge 1$ is the total number of scheduling times slots. For simplicity, we assume that the duration of each slot is normalized to a unit time; as a result, power and energy are used interchangeably in this paper. As shown in Fig. \ref{fig:SystemModel}, we consider $M \ge 1$ number of users, indexed by $m$, $m\in {\cal M}=\{1,\cdots,M\}$.  Particularly, each user can be a single energy consumer (residential, commercial, and/or industrial) or a group of consumers controlled by an  aggregator.  We assume that users have their individual  renewable energy generators; thus, can supply a part  or all of their load over time. However, a single large-scale energy storage system (ESS) is shared among all users, where they can  charge/discharge to/from it whenever necessary. 
Moreover, users are all connected to the main grid, which consists of conventional fossil fuel based energy generation units, and can draw energy from it in case of renewable energy deficit.  
We assume that a central controller who is trusting of all users optimizes the amount of  charging/discharging to/from the shared ESS while ensuring that  practical constraints of the system are satisfied. 

Let $0 \le C_{mn} \le \overline{C}$ and $0 \le D_{mn} \le \overline{D}$ denote the energy charged/discharged to/from the shared ESS by  user $m$ at time slot $n$, respectively, where $\overline{C}>0$ and $\overline{D}>0$ are the maximum charging and discharging rates of the shared ESS, respectively. Furthermore, there are some energy losses during the charging and discharging processes of the ESS in practice, which are specified by charging and discharging efficiency parameters, denoted by $0< \dot{\alpha} < 1$ and $0 <\ddot{\alpha} < 1$, respectively. Let $S_n\ge 0$ denote the available energy in the shared ESS at the beginning of time slot $n$,  which can be derived recursively as follows:
\begin{align}
S_{n+1}=S_n+\dot{\alpha} \sum_{m=1}^{M}{C_{mn}} - \frac{1}{\ddot{\alpha}} \sum_{m=1}^{M}{D_{mn}} .
\end{align}
In addition, a practical ESS has finite capacity and cannot be fully discharged to avoid deep charging. We thus have the following constraints for the states of the shared ESS over time
\begin{align}\label{eq:storage}
\underline{S} \le S_{n} \le \overline{S},~\forall n \in {\cal N}, 
\end{align}
where $\underline{S} \ge 0$ and $\overline{S} \ge 0$ are the minimum and maximum   allowed states of the shared ESS, respectively. By default, we assume $\underline{S} \le S_{1} \le \overline{S}$.

Denote $R_{mn}\ge 0$ and $L_{mn}\ge 0$ as the renewable energy generation and the load of user $m$ at time slot $n$. We then define $\Delta_{mn}=R_{mn}-L_{mn}$ as the {\it net energy profile} of user $m$. Note that $\Delta_{mn}$ can be either positive (case of surplus renewable energy) or negative (case of renewable energy  deficit). In general, $\Delta_{mn}$'s are stochastic due to the randomness in both the renewable energy generation and load, but can be predicted with finite errors. However, in this paper, we assume that $\Delta_{mn}$'s are perfectly predicted and thus are known to the users prior to the scheduling, e.g., day-ahead energy management.\footnote{In the practical case of stochastic net energy profiles, we can  design  online algorithms for the real-time implementation using, e.g., sliding-window technique given in  \cite{Katayoun_journal}. Due to the space limitation,  we do not discuss the online energy management in this paper.} By denoting $G_{mn} \ge 0$ as the  energy drawn from the main grid by user $m$ at time slot $n$, we have the following constraints for each user $m$:
\begin{align}\label{eq:Nut}
G_{mn}-C_{mn}+D_{mn}+\Delta_{mn} \ge 0,~\forall n \in {\cal N}.
\end{align}
The constraints in (\ref{eq:Nut}) ensure that the load of user $m$ is satisfied in all time slots,  from its own renewable energy generation, the shared ESS, and/or the main grid. Drawing energy from the main grid incurs some cost for user $m$, which can be modelled by a set of time-varying functions $f_{mn}(G_{mn})$, $n=1,\ldots,N$. Specifically, we assume that each  $f_{mn}(G_{mn})$ is convex \cite{Boyd} and increasing over $G_{mn}$, e.g., piece-wise linear and/or quadratic functions \cite{Wood}.

%**********************************
\section{Problem Formulation: Shared ESS}\label{Sec:Shared_ESS} \label{sec:PF-Shared}
As discussed earlier, drawing energy from the main grid  results in some cost for individual users. As a result, each user wants to use the shared ESS as much as possible in order to maximize its own {{\it profit}, which is  defined as the difference between its energy costs without and with the shared ESS. In this case, the central controller should ensure that the  ESS is shared appropriately (fairly) among all users while satisfying the practical constraints of the system.  
In the following, we first formulate the profit maximization problem for each individual user, and derive its profit as a function of the charging/discharging values. Next, we apply the technique of profit coefficients under which the central controller proportionally allocates the total profit to users.
 
Given $\{ G_{kn} \ge 0\}_{k\neq m}$, $\{0 \le C_{kn} \le \overline{C}\}_{k\neq m}$, $\{ 0 \le D_{kn} \le \overline{D}\}_{k\neq m}$, satisfying constraints in (2) and (3), $\forall k \neq m$, the  {\it profit maximization problem} for each user $m$ is given by  
\begin{align}
\mathrm{(P1)}\hspace{-.7mm}:&~\hspace{-1mm}\mathop{\mathtt{max}}_{\{G_{mn}\}_{n \in \cal N},\{C_{mn}\}_{n \in \cal N},\{D_{mn}\}_{n \in \cal N}}
~\hspace{-1mm}{\hat{f}_m\hspace{-.5mm}-\hspace{-.5mm}\sum_{n=1}^{N}{f_{mn}(G_{mn})} }\nonumber \\ 
\mathtt{s.t.}
&~ \underline{S} \le S_n \le \overline{S},~\forall n \in {\cal N} \nonumber\\
&~ G_{mn}-C_{mn}+D_{mn}+\Delta_{mn} \ge 0,~\forall n \in {\cal N} \nonumber\\
&~ G_{mn}\hspace{-.4mm} \ge\hspace{-.4mm} 0,\hspace{-.7mm}~0\hspace{-.5mm}\le\hspace{-.4mm} C_{mn} \hspace{-.4mm}\le \overline{C},\hspace{-.7mm}~\hspace{-.5mm}0\hspace{-.4mm}\le\hspace{-.4mm} D_{mn} \hspace{-.4mm}\le \overline{D},\hspace{-.7mm}~\hspace{-.4mm}\forall n \in {\cal N}, \nonumber
\end{align}
where $\hat{f}_m=\sum_{n=1}^{N}f_{mn}([-\Delta_{mn}]^+)$, with $[x]^+=\max\{0,x\}$, denotes the energy cost for user $m$ when the shared ESS is not deployed. In (P1), we can first solve the problem over $\{G_{mn}\}_{n\in \cal N}$, given fixed $\{0\le C_{mn}\le \overline{C}\}_{n\in \cal N}$ and $\{0 \le D_{mn} \le \overline{D}\}_{n\in \cal N}$. By solving the resulted problem, we obtain  $G_{mn}=[C_{mn}-D_{mn}-\Delta_{mn}]^+$, $\forall n \in \cal N$. Accordingly, the user's profit can be expressed as a function of its charging/discharging values as follow:
\begin{align}\label{eq:Profit}
&P_m(\{C_{mn}\}_{n\in \cal N},\{D_{mn}\}_{n\in \cal N})=
\nonumber\\&\sum_{n=1}^{N}{\hspace{-.5mm}f_{mn}([-\Delta_{mn}]^+\hspace{-.3mm})}\hspace{-.8mm}-\hspace{-.8mm}\sum_{n=1}^{N}{\hspace{-.5mm}f_{mn}([C_{mn}-D_{mn}-\Delta_{mn}]^+}\hspace{-.3mm}).
\end{align} 

Next, the central controller needs to jointly design $\{C_{mn}\}^{m\in \cal M}_{n\in \cal N},\{D_{mn}\}^{m\in \cal M}_{n\in \cal N}$ for all users such that their individual  profits are maximized proportionally while  practical constraints of the shared ESS are satisfied. Let $P_m(\{C_{mn}\}_{n\in \cal N},\{D_{mn}\}_{n\in \cal N})=\beta_m t^*$, where $0 \le \beta_m \le 1$, $\forall m \in {\cal M}$, satisfying $\sum_{m=1}^{M}{\beta_m}=1$, are given  profit coefficients\footnote{In practice, there are different approaches to design  the profit coefficients $\beta_m$'s. For instance, consider the scenario that users invest to purchase a bulk battery, according to their budget. In this case, the system operator can set $\beta_m$'s such that users benefit from the shared ESS according to their initial investment. As another example, consider a scenario that the  ESS is already installed in the system, e.g., by the government funding. In this case, the central controller sets the use of ESS such that users with higher average renewable energy deficit can draw more from the shared ESS.} that specify the proportion of the total profit received by individual users, and $t^* \ge 0$ denotes the maximum total profit of all users, i.e., $t^*=\sum_{m=1}^{M}P_m(\{C_{mn}\}_{n\in \cal N},\{D_{mn}\}_{n\in \cal N})$.
Specifically, $t^*$ can be obtained by solving the following optimization problem:
\begin{align}
&\mathrm{(P2)}:~\mathop{\mathtt{max}}_{t, \hspace{.2mm}\{C_{mn}\}^{m\in \cal M}_{n\in \cal N},\{D_{mn}\}^{m\in \cal M}_{n\in \cal N}}~{t}\nonumber \\
\mathtt{s.t.}
&~ \underline{S} \le S_n \le \overline{S},~\forall n \in {\cal N},\nonumber\\
&~ P_m(\{C_{mn}\}_{n\in \cal N},\{D_{mn}\}_{n\in \cal N}) \ge \beta_m t,~ \forall m \in {\cal M} \nonumber\\
%&~ C_{mn} \ge 0,~D_{mn} \ge 0,~ t\ge0, ~\forall m \in {\cal M}, ~\forall n \in {\cal N}.
&~\hspace{-.4mm}~0\hspace{-.4mm}\le\hspace{-.4mm} C_{mn} \hspace{-.4mm}\le \overline{C},\hspace{-.4mm}~\hspace{-.4mm}0\hspace{-.4mm}\le\hspace{-.4mm} D_{mn} \hspace{-.4mm}\le \overline{D},\hspace{-.6mm}~ t\ge0, \hspace{-.6mm}~\forall m \hspace{-.4mm}\in \hspace{-.4mm}{\cal M}, \hspace{-.6mm}~\forall n\hspace{-.4mm} \in\hspace{-.4mm} {\cal N}\hspace{-.4mm}.\nonumber
\end{align} 
It can be easily verified that (P2) is  convex \cite{Boyd}, and thus can be solved using  standard convex optimization techniques such as interior point method. Alternatively, (P2) can be solved by a bisection search over $t$, where in each search iteration, it suffices to solve a feasibility problem that checks whether all constraints of (P2) can be satisfied for given $t$. Eventually, $t$ converges to its optimal value that is $t^*$. In this paper, we use CVX software\cite{CVX} to derive the optimal solution to (P2), i.e., $t^*$,  $\{C_{mn}^*\}^{m\in \cal M}_{n\in \cal N}$, and $\{D_{mn}^*\}^{m\in \cal M}_{n\in \cal N}$.
%**********************************
\section{Performance Benchmark: Distributed ESSs}\label{Sec:Benchmark}
In this section, we consider a system setup of distributed ESSs, where users own their individual small-scale ESSs, but do not share energy  with each other.
\begin{figure}[t!]
\centering
\includegraphics[width=7.6cm]{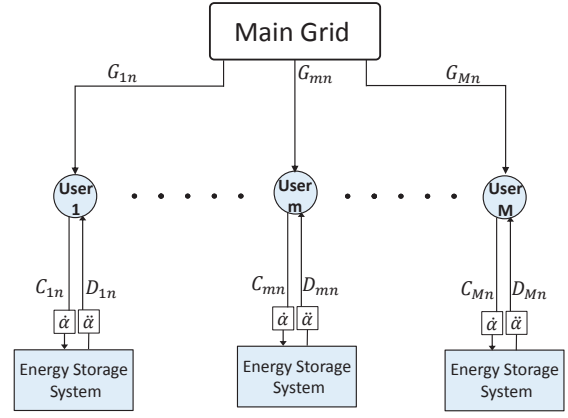}\\ 
\caption{Case of distributed small-scale ESSs.}\label{fig:SystemModel_Split}\vspace{-4mm}
\end{figure}
In this case, the state of the ESS for each user $m$, denoted by $S_{mn}$ at time slot $n$,  can be derived as
\begin{align}
S_{mn+1}=S_{mn}+\dot{\alpha}_m{C_{mn}} - \frac{1}{\ddot{\alpha}d_m} {D_{mn}}, \label{Storage_Split}
\end{align}
where $ 0<\dot{\alpha}_m<1$ and $0<\ddot{\alpha}_m<1$ are charging and discharging efficiency parameters, respectively. We then have the following constraints for the ESS of user $m$:
\begin{align}\label{eq:storage_distributed}
\underline{S}_m \le S_{mn} \le \overline{S}_m,~\forall n \in {\cal N}, 
\end{align}
where $\underline{S}_m\ge 0$ and $\overline{S}_m\ge 0$ are the minimum and maximum allowed states of the ESS. The charging and discharging values should also satisfy $0 \le C_{mn} \le \overline{C}_m$ and $0 \le D_{mn} \le \overline{D}_m$, where  $\overline{C}_m>0$ and $\overline{D}_m>0$ are the maximum charging and discharging rates, respectively. To have a fair comparison with the case of shared ESS  in Section \ref{sec:PF-Shared},  we set $\underline{S}=\sum_{m=1}^{M}\underline{S}_m$, $\overline{S}=\sum_{m=1}^{M}\overline{S}_m$, $\overline{C}=\sum_{m=1}^{M}\overline{C}_m$, and $\overline{D}=\sum_{m=1}^{M}\overline{D}_m$. 

We now proceed to formulate the profit maximization problem for each user $m$. This problem  can be formulated for each  user independently, since  ESS constraints are not coupled over users any more. 
Hence, the profit maximization  problem for each user $m$ can be given as follows:
\begin{align}
\mathrm{(P3)}&:~\hspace{-1mm}\mathop{\mathtt{max}}_{\{G_{mn}\}_{n\in \cal N},\{C_{mn}\}_{n\in \cal N},\{D_{mn}\}_{n\in{\cal N}}}~\hspace{-1mm}
{\hat{f}_m-\sum_{n=1}^{N}{f_{mn}(G_{mn})} }\nonumber \\ 
\mathtt{s.t.}
&~ \underline{S}_m \le S_{mn} \le \overline{S}_m,~\forall n \in {\cal N} \nonumber\\
&~ G_{mn}-C_{mn}+D_{mn}+\Delta_{mn} \ge 0,~\forall n \in {\cal N} \nonumber\\
&~ G_{mn}\hspace{-.4mm} \ge\hspace{-.4mm} 0,\hspace{-.7mm}~0\hspace{-.5mm}\le\hspace{-.4mm} C_{mn} \hspace{-.4mm}\le \overline{C}_m,\hspace{-.7mm}~\hspace{-.5mm}0\hspace{-.4mm}\le\hspace{-.4mm} D_{mn} \hspace{-.4mm}\le \overline{D}_m,\hspace{-.7mm}~\hspace{-.4mm}\forall n \in {\cal N}.\nonumber
\end{align}
It can be easily verified that (P3) is  convex \cite{Boyd}, and thus can be solved using CVX software\cite{CVX}.  

In the next section, we present a numerical example to reveal the possible profit gain resulting from the shared ESS, compared to the case of distributed ESSs. We also discuss the effectiveness of the shared ESS in users' profit maximization under different system settings.
%**********************************
\section{Simulation Results}\label{Sec:Simulations}
\begin{figure}[t!]
	\centering
	\includegraphics[width=6.5cm]{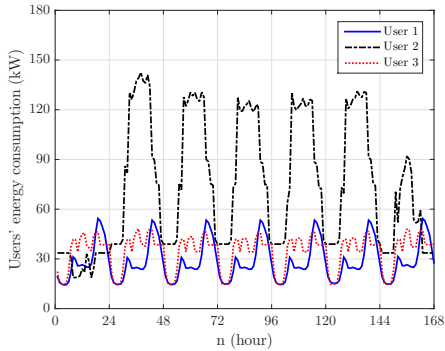}\\ 
	\caption{Load profiles of the three users over one week.}\label{fig:Users_Load_Profile}
\end{figure}
\begin{figure}[t!]
	\centering
	\includegraphics[width=6.5cm]{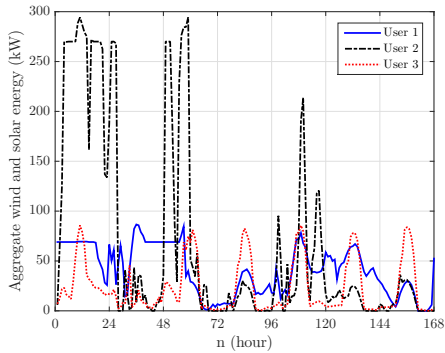}\\ 
	\caption{Renewable energy generation (both solar and wind) of the three users over one week.}\label{fig:Users_RE}
\end{figure}
In this section, we consider a system with three users $M=3$ based on the real data available from California,
US, over one week $N=168$ (from 1 January, 2006 to 7 January, 2006) \cite{Load_Profile,NREL,NREL_Solar,Price_CAISO}. Specifically, user 1 is modelled as an apartment of $30$ units, user 2 as a medium-size office, and user 3 as a restaurant \cite{Load_Profile}. 
The hourly energy consumption of the three users is shown in Fig.  \ref{fig:Users_Load_Profile}.
%, from which it is observed that users' load profiles are quite different over time.
%
We assume that each user has its own renewable energy generators, including both solar and wind, with generation profiles shown in Fig. \ref{fig:Users_RE} \cite{NREL,NREL_Solar}. 
For the shared ESS setup, we set  $\dot{\alpha}=0.7$, $\ddot{\alpha}=0.8$, $\underline{S}=0$,  and $\overline{S}= 1$ MW.  On the other hand, for the distributed ESSs setup, we set charging and discharging efficiency parameter of individual ESSs as $\dot{\alpha}_m=0.7$, $\ddot{\alpha}_m=0.8$, $\forall m\in \cal M$, respectively. We also set   $\underline{S}_m=0$, $\forall m\in \cal M$,   $\overline{S}_{1}=300$ kW,  $\overline{S}_{2}=600$ kW, and $\overline{S}_{3}=100$ kW. %(or $\overline{S}_{1}=450$ kW,  $\overline{S}_{2}=950$ kW, and $\overline{S}_{3}=150$ kW). 
%where $\overline{S}_{1}+\overline{S}_{2}+\overline{S}_{3}= \overline{S}$.
We model the cost function of purchasing energy from the main grid  as $f_{mn}=\lambda_{mn}G_{mn}$, where we set $\lambda_{mn}=45$ $\$$/kW,  $\forall m\in {\cal M}$, $\forall n\in {\cal N}$  \cite{Price_CAISO}. 
%
%**********************************
\subsection{Shared versus Distributed ESSs}
\begin{table}[t!]
	\centering
	\caption{Comparison between the shared and distributed ESSs}
	\label{Table}
	\begin{tabular}{|c|c|c|c|}
		\hline
		\multicolumn{2}{|c|}{ESS values}                                                                                                        & \multicolumn{2}{c|}{Total profit gain of all users} \\ \hline
		Shared ESS        & Distributed ESS                                                                                                     & Shared ESS       & Distributed ESS      \\ \hline
		$\overline{S}=1$ MW   & \begin{tabular}[c]{@{}c@{}}$\overline{S}_{1}=300$ kW\\ $\overline{S}_{2}=600$ kW\\ $\overline{S}_{3}=100$ kW\end{tabular} & $28.4$$\%$       & $18.5$$\%$           \\ \hline
		$\overline{S}=1.5$ MW & \begin{tabular}[c]{@{}c@{}}$\overline{S}_{1}=450$ kW\\ $\overline{S}_{2}=900$ kW\\ $\overline{S}_{3}=150$ kW\end{tabular} & $32.6$$\%$       & $25.4$$\%$           \\ \hline
	\end{tabular}
\end{table}
First, we set the profit coefficients as $\beta_1=0.3$, $\beta_2=0.6$, and $\beta_3=0.1$. 
%Herein, we aim to show that the shared ESS can help increase the total profit compared to the case of distributed ESSs. 
Table \ref{Table} shows the total profit gain of all users, i.e., energy cost saving due to using ESS, resulting from the two system setups of shared and distributed ESSs. It is observed that shared ESS can  result in higher profit, e.g., profit gain with a shared ESS of  $\overline{S}=1$ MW is $28.4\%$, while distributed ESSs setup with $\overline{S}_{1}=300$ kW,  $\overline{S}_{2}=600$ kW, and $\overline{S}_{3}=100$ kW yields a lower profit gain of $18.5\%$. This is because  the surplus energy of one user can be utilized by others with energy deficit and users can thus cooperatively exploit the shared ESS. In addition, the shared ESS can avoid renewable energy curtailments more effectively over the case of distributed ESSs, due to its higher capacity compared to each distributed ESS.  
\begin{figure}[t!]
	\centering
	\includegraphics[width=6.7cm]{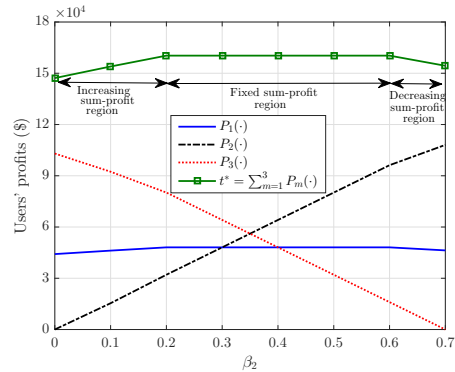}\\ 
	\caption{Profit of users over profit coefficient $\beta_2$, given $\beta_1=0.3$ and $\beta_3=0.7-\beta_2$.}\label{fig:Profit_Over_Beta}
\end{figure}
%**********************************
\subsection{Impact of Profit Coefficients}
As discussed earlier, profit coefficients  $\beta_m$'s are set by the system operator to allocate a certain proportion  of the total profit to each user. Fig. \ref{fig:Profit_Over_Beta} shows the individual profits of users, i.e., $P_m(\cdot)$'s given in (\ref{eq:Profit}), versus $\beta_2$, with $\beta_1=0.3$ and $\beta_3=0.7-\beta_2$. It is observed that $P_1(\cdot)$ remains almost constant over $\beta_2$, since $\beta_1$  is assumed to be  fixed in this simulation. In addition, $P_2(\cdot)$ increases over $\beta_2$, since higher proportion of the total profit is allocated to user 2 as $\beta_2$ increases, while the opposite is true for $P_3(\cdot)$.
%
%
%**********************************
\subsection{Impact of Renewable Energy Diversity}
Herein, by keeping the load profile of each user unchanged, we consider a scenario that all users  have only solar energy generators (low diversity), which is in contrast to the initial setup of solar together with wind energy generators (high diversity). The  solar energy generation of users is shown in Fig. \ref{fig:User_RE_Only_PV}. 
The total profit gain of all users for the two cases of high and low renewable energy diversity are shown in Fig. \ref{fig:Shared_ESS_Evaluation}. It is observed that the case with high diversity of renewable energy generators  yields a higher profit gain compared to the low diversity case. This is because when diversity is high, it is  more likely that the  energy surplus/deficit in users' renewable energy profiles do not happen concurrently; thus, the  surplus energy in one user can compensate the energy deficit in others. In addition, it is observed that the profit gain is almost constant for $\overline{S} > 1$ MW in the low diversity case, while in the high diversity case, it happens for $\overline{S} >5$ MW. This is because when the diversity is low, the chance for energy cooperation among users using the shared ESS is small.  
%ntegrating various number of  solar and wind energy generators in user level yields more diverse net energy profiles and 
%
%
\section{Conclusion and Future Work}\label{Sec:Conclusion}
In this paper, we study the shared ESS  management problem for users with renewable energy integration. We propose an algorithm under which a central controller optimizes the charging/discharging power to/from the shared ESS by all users, given a profit coefficient set. 
By comparing the performance of this setup in terms of profit gain with the conventional setup of distributed small-scale ESSs, we show that the shared ESS  can potentially increase the total profit of all users (up to $10\%$ in our simulations). Our results are helpful in practical systems  where installing individual ESS for each user is either very costly (due to the large number of users) or requires space that is not available. 
Devising efficient online algorithms for the real-time energy management of the shared ESS, given stochastic renewable energy generation/load, is an interesting future direction. %It is also important to consider  operational costs in both shared and distributed ESSs systems for a more practical  evaluation. 
\begin{figure}[t!]
	\centering
	\includegraphics[width=6.5cm]{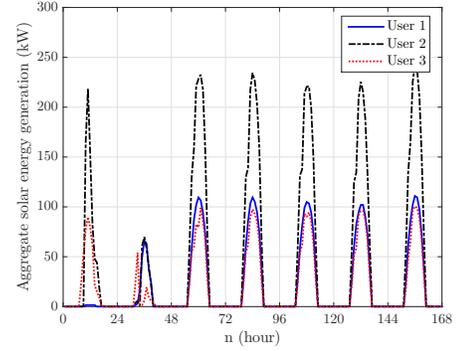}\\ 
	\caption{Renewable energy generation (only solar) of the three users over one week.}\label{fig:User_RE_Only_PV} \vspace{-3mm}
\end{figure}
\begin{figure}[t!]
	\centering
	\includegraphics[width=6.7cm]{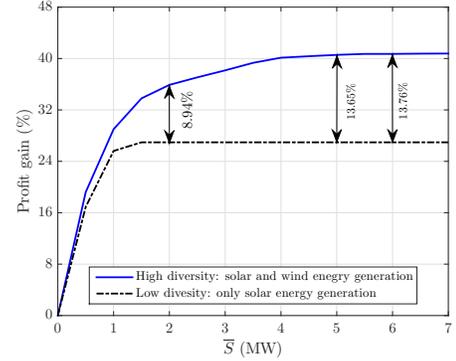}\\ 
	\caption{Impact of renewable energy  diversity on the performance of shared ESS.}\label{fig:Shared_ESS_Evaluation}\vspace{-3mm}
\end{figure} 
%**********************************Shared ESS performance evaluation in two system setups of only solar energy generators and solar and wind energy generators.

\end{document}